\documentclass[authoryear,preprint]{elsarticle}
\usepackage{graphicx}
\usepackage{lscape}
\usepackage{amssymb}
\usepackage{natbib}

\journal{Chemical Physics}

\begin{document}

\begin{frontmatter}

\title{Visible photodissociation spectroscopy of PAH cations and 
derivatives in the PIRENEA experiment}

\author[label1,label2]{F. Useli-Bacchitta}
\author[label1,label2]{A. Bonnamy}
\author[label3]{G. Mulas}
\author[label3]{G. Malloci}
\author[label1,label2]{D. Toublanc}
\author[label1,label2]{C. Joblin \corref{cor}}
\cortext[cor]{Corresponding author. Tel: +33\textendash05 61 55 86 01;
Fax: +33\textendash05 61 55 67 01}
\ead{christine.joblin@cesr.fr}

\address[label1]{Universit\'{e} de Toulouse,  UPS, CESR,
9 avenue du colonel Roche, F-31028 Toulouse cedex~4, France}
\address[label2]{CNRS, UMR5187, F-31028 Toulouse, France}
\address[label3]{Istituto Nazionale di Astrofisica - Osservatorio 
Astronomico di Cagliari, Strada n. 54, Loc. Poggio dei Pini, 
09012 Capoterra (CA), Italy}

\begin{abstract}
The electronic spectra of gas-phase cationic polycyclic aromatic 
hydrocarbons (PAHs), trapped in the Fourier Transform Ion Cyclotron 
Resonance cell of the PIRENEA experiment, have been measured 
by multiphoton dissociation spectroscopy in the 430--480~nm 
spectral range using the radiation of a mid-band optical parametric oscillator laser. 
We present here the spectra recorded for different species 
of increasing size, namely the pyrene cation (${C_{16}H_{10}}^+$), 
 the 1\textendash methylpyrene cation (${CH_{3}-C_{16}H_{9}}^+$), 
the coronene cation (${C_{24}H_{12}}^+$), 
and its dehydrogenated derivative ${C_{24}H_{10}}^+$. 
The experimental results are interpreted with 
the help of time-dependent density functional theory calculations and analysed using spectral information on the 
same species obtained from matrix isolation spectroscopy data. A kinetic 
Monte Carlo code has also been used, in the case of pyrene and coronene 
cations, to estimate the absorption cross-sections of the measured electronic 
transitions. Gas-phase spectra of highly reactive 
species such as dehydrogenated PAH cations are reported for the first time.

\end{abstract}


\end{frontmatter}

\section{Introduction}
\label{Intr}

The measurement of the electronic spectra of large molecular ions in the 
gas phase still remains a significant experimental challenge. Typical 
problems include the production in the gas-phase of these ions, 
preferentially at low internal temperature, the difficulty of creating 
sufficient ion densities to apply standard spectroscopic techniques and of
isolating these species that are often unsaturated 
and highly reactive. Polycyclic aromatic hydrocarbons (PAHs) provide a 
good case to illustrate these problems. Astrophysicists are particularly 
interested in these molecules as they are believed to be the major carriers 
of the so-called aromatic infrared bands (AIBs)~\cite{Leg84,All85,Sell}, 
a set of discrete emission bands dominating the mid-infrared spectra of 
many galactic and extragalactic objects~\cite{Tie05}. PAHs could contain 
a large fraction of the carbon present in the interstellar medium (typically 20\%~\cite{Job92}).
Neutral and ionised PAHs are also considered as attractive candidates for the Diffuse Interstellar 
Bands (DIBs), some discrete absorption features observed throughout the 
visible and near-infrared spectral range, whose origin is still investigated 
\cite{Leg85,Herb,van}. The possible assignment of PAH cations as carriers of 
some DIBs has motivated, during the past two decades, an extensive laboratory 
effort to find the most suitable techniques to record the electronic spectra 
of these species under conditions similar to those found in space. Most 
laboratory information has been obtained so far from matrix isolation 
spectroscopy (MIS)~\cite{Sal91,Szcsep,Ruit}. One major limitation of this 
technique, however, is the difficulty { to know the identity of the species} that are produced 
in the matrix upon UV irradiation of the neutral precursor. In the gas-phase, 
various methods have been successfully applied like cavity ring down 
spectroscopy (CRDS) in free jets~\cite{Rom99,Suk} and photodissociation 
spectroscopy of weakly bound ionic complexes~\cite{Brec,Boud}. Another 
interesting approach is the coupling of ion traps, where ions can be easily 
collected, mass-selected and thermalised, with multiphoton dissociation (MPD) 
spectroscopy. This approach is extensively used, for instance, in the infrared 
(IR) on PAHs~\cite{Oom2000,Oom01} and on cationic iron-PAH complexes 
\cite{Szczep06,Simon}. In the visible spectral range, Rolland et 
al. have performed MPD of phenanthrene and anthracene cations~\cite{Roll}. 
An important advantage of this approach is that parent ions and photofragments 
can be mass-selected, thus removing any ambiguity on the identity of the recorded 
species, with the exception of simultaneously produced isomers which are not 
distinguishable with mass spectrometry.

In this work we report the visible MPD spectra of gas-phase cationic 
PAHs which have been produced and isolated in the PIRENEA Fourier Transform 
Ion Cyclotron Resonance (FT-ICR) ion trap mass spectrometer~\cite{Job02},
and irradiated with a mid-band OPO tunable laser. The 
purpose of this study is to obtain gas-phase spectroscopic data on different 
isolated, ionised PAHs and their derivatives, that can be useful for the 
pre-selection of the most promising candidates for some of the DIBs. We have investigated, in particular, the 430-480~nm spectral range in which the 
strongest optical DIB, the 4430 \AA~band, has been observed. { The comparison of 
our measurements with previous matrix and gas-phase data, when available, allowed 
us to assess the validity of our experimental method in obtaining the band positions. 
The band profile obtained with this technique, however, is not that of cold ions, which is a necessary condition to match the experimental spectra to the astronomical ones and definitively identify DIB carriers. That is why the technique applied here can only be used to make a pre-selection of possible DIB candidates. The reported measurements}
include the first results obtained in the gas-phase for species containing more 
than 20 carbon atoms, e.g.~coronene cation, C$_{24}$H$_{12}^+$ and its
dehydrogenated species ${C_{24}H_{10}}^+$, { recently proposed as possible 
candidates for some of the DIBs~\cite{Duley}.} 

The experimental set-up is described in Sect.~\ref{Met}. 
Our results are discussed in the following Sect.~\ref{Res}, 
where they are compared to neon-matrix spectra and complemented with 
time-dependent density functional theory (TD-DFT) calculations. In Sect.~\ref{Model} 
a kinetic Monte Carlo code has been used to model the photophysics of the irradiated 
species and to derive the absorption cross-section for two of the studied species. 
The conclusions are finally presented in Sect.~\ref{Con}.

\section{Methods}
\label{Met}
\subsection{Experimental methods}

The PIRENEA set-up, initially described by Joblin et al.~\cite{Job02}, was used 
to produce and trap the cationic species under study. PIRENEA (Piege \`a 
Ions pour la Recherche et l'Etude de Nouvelles Especes Astrochimiques) is a 
home-built Fourier Transform Ion Cyclotron 
Resonance (FT-ICR) mass spectrometer with the additional characteristics of a 
cold environment generated by a set of cryogenic shields ($T_{trap} \sim 35 K$), 
and of low pressure ($P~\leq 10^{-10}$ mbar) that are of interest to approach the physical 
conditions of interstellar space. Desorption and ionisation of molecular species result 
from laser irradiation of a solid target using the fourth harmonic, 
$\lambda$=266~nm, of a Nd:YAG laser (Minilite II, Continuum). Trapping of 
ionised species is achieved through the conjugated action of an axial magnetic 
field, generated by a superconducting magnet of 5~T, and static electrical 
potentials. {The magnetic field confines the ions in the radial direction while the static electric field traps them in the axial direction.} Fragment species (such as dehydrogenated species) can be produced \emph{in situ} by irradiation of the parent ion with a 150~W Xe arc lamp focused by an elliptical
mirror at the centre of the cell. The ions of interest can then be selected and
isolated in the ICR cell by selective ejection of all the other species~\cite{Mars98}. 
In this process, however, the ion cloud is perturbed to some extent making it necessary to inject buffer gas (He) in order to relax it. 
{The buffer gas, injected to act as a medium 
to enable the ions to exchange kinetic energy with the cold walls of the trap so that the cloud 
can cool down, mediates, as a side effect, the exchange 
of energy among translational and internal degrees of freedom of the molecules. Therefore if the ion cloud has a large residual energy, this can populate significantly some vibrational levels of the ions.}

The opened cylindrical geometry of the cell, designed with optical 
experiments in mind, provides wide access to the centre of the trap where ions 
are confined. Direct absorption spectroscopy is however not possible due to the
low quantity of trapped species (ion cloud containing typically $10^7$ ions). 
An action spectrum is therefore recorded which gives the efficiency of 
fragmentation as a function of wavelength. We performed MPD spectroscopy for 
different PAH cations and derivatives using a mid-band tunable optical parametric oscillator (OPO) laser
system (Panther EX, Continuum, 5~$cm^{-1}$ bandwidth, 5~ns pulse duration) operating at 
10~Hz, to induce fragmentation of the species in the 430--480~nm spectral range. 
Whenever the laser wavelength is in resonance with an 
electronic transition of the cation, the sequential absorption of multiple photons can 
take place and the dissociation can proceed. After each photon absorption event,
the energy absorbed in an electronic excited state is rapidly converted  into 
vibrational energy in the ground state, as generally observed for PAHs~\cite{Leg89}.
Absorption of photons will occur from a distribution of vibrational states whose 
temperature increases with the number of absorbed photons. Therefore, even when the ions to be 
probed are cold before the beginning of the MPD scan, they are hot after absorbing 
the first of the series of photons needed to dissociate them.

\subsection{Computational details}
To help the interpretation of the measured spectra we computed the energies 
and intensities for vertical transitions to the low-lying electronic states of 
the species under investigation. We used the density functional theory (DFT) 
\cite{hoh64,koh64,jon89} and its time-dependent extension (TD-DFT) 
\cite{run84,cas95,bau96}, which are the methods of choice for such large
molecules. We used the \textsc{TURBOMOLE V6.0.1} package \cite{Ahl} making 
use of the resolution of identity approximation for computing the electronic 
Coulomb interaction \cite{Eich}. This approach is based on the expansion 
of molecular electron densities in a set of atom-centered auxiliary basis sets
leading to expressions involving three-center electron repulsion integrals; 
this usually leads to a more than tenfold speedup compared to the conventional 
method based on four-center electron repulsion integrals. 

The calculation of the electronic absorption spectra requires the previous 
knowledge of the ground-state optimised geometries. In all cases, with the 
exception of coronene cation (see section \ref{Res}), no symmetry 
constraints were adopted by assuming the C$_1$ point group. 
The electronic ground-state were found to be the doublet for all
species considered. Based on calibration calculations performed for small PAHs, 
we used the split valence polarisation (SVP) basis set \cite{svp} in conjunction with the BP86 
exchange-correlation functional, a combination of the Becke's 1988 exchange 
functional \cite{becke}, and the Vosko-Wilk-Nusair \cite{vosko}
and Perdew's 1986 \cite{perdew} correlation functionals. Despite the smallness
of the basis set, the computed transition energies in the spectral range of 
interest (430-480~nm) are found to be systematically in closer agreement with 
experiments as compared to results obtained with the larger basis TZVP of triple 
zeta valence quality \cite{tzvp}.
{Typical errors in the computed values are of the order of a few tenths of eV.} 

\section{Results and discussion}
\label{Res}

The spectra were recorded monitoring the relative fragmentation yield of the 
species (ratio between the photofragment abundance and the total abundance of 
ions, parent ion plus photofragments) as a function of the OPO laser 
wavelength. { All the dissociation mass-spectra exhibited the same main fragmentation 
channel resulting in the sequential loss of hydrogen atoms. The 1\textendash methylpyrene 
cation also exhibits the loss of $C_{2}H_{2}$.} We scanned with the laser the spectral range between 430 and 
480~nm, with a 2 nm scan step, recording, at each wavelength, 10 mass spectra 
and then averaging them to give the points represented in the plots. The laser energy 
is kept constant during the wavelength scan. 
The measured spectra are discussed separately below.
The energies of the relevant computed transitions are marked by vertical bars, with 
an height proportional to the computed oscillator strengths,
and superimposed on the experimental spectra. 

\begin{figure}
\begin{center}
\includegraphics[scale=.15,clip,trim=60 60 60 60]{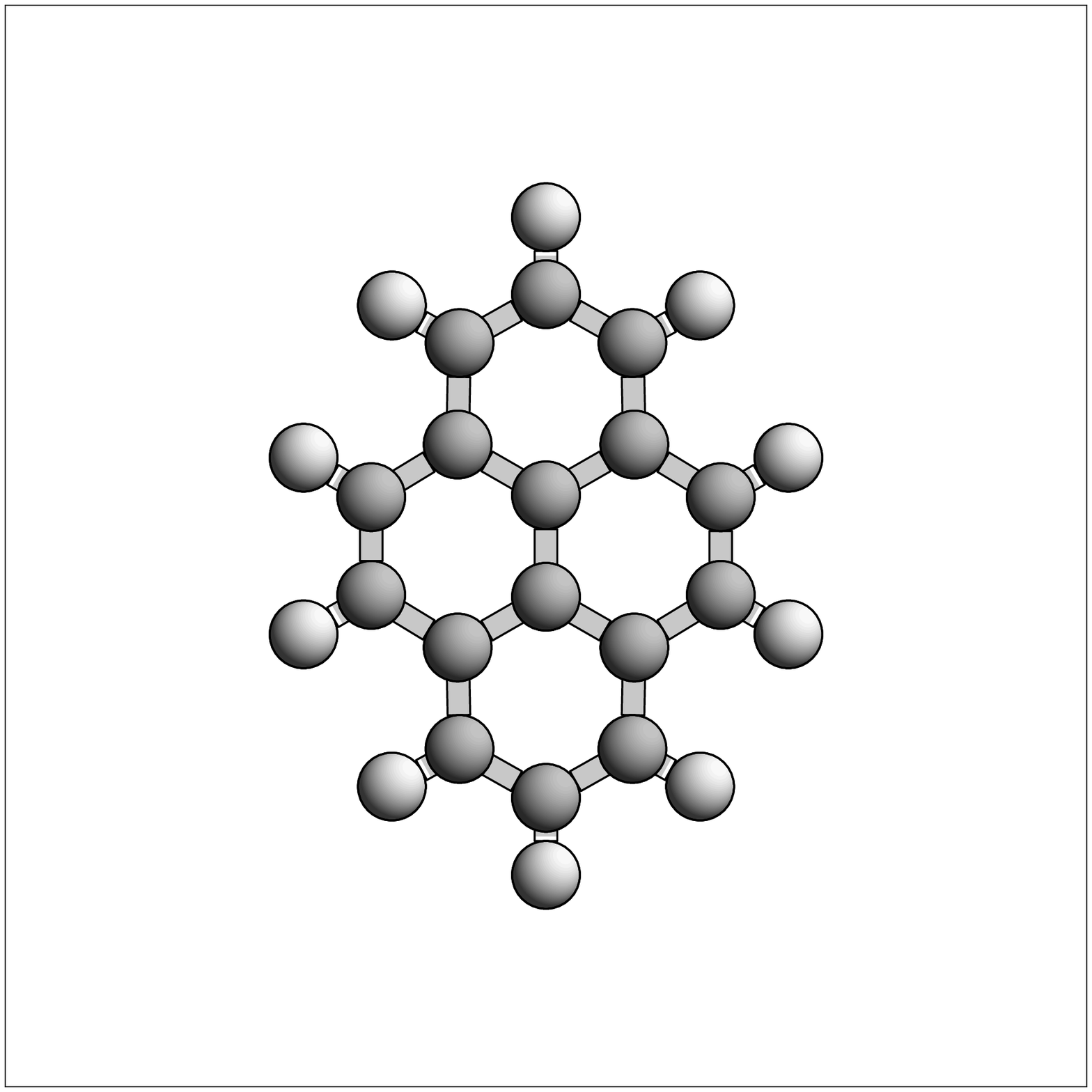} \hspace{-8mm}(a) \hspace{5mm}
\includegraphics[scale=.15,clip,trim=60 60 60 60]{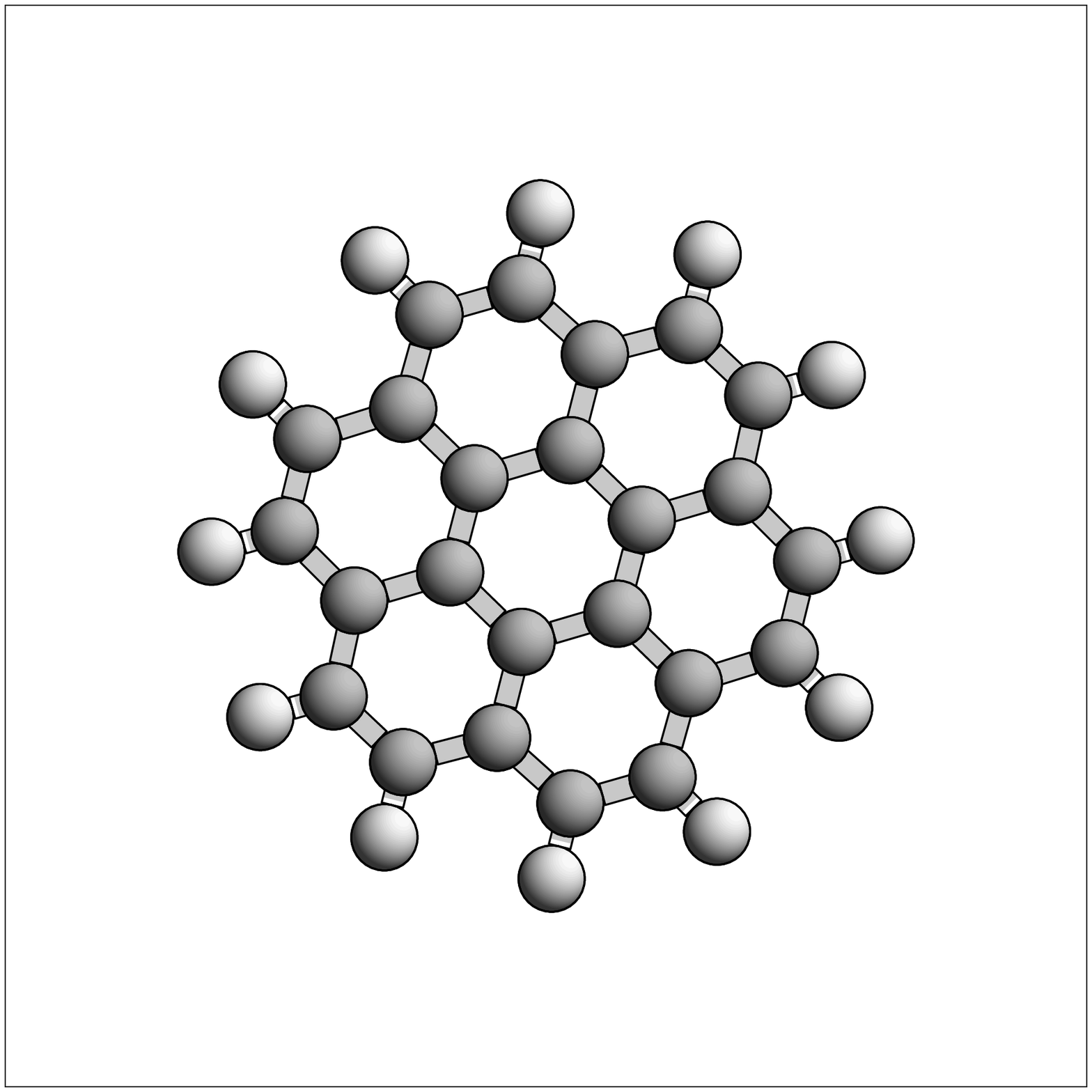} \hspace{-6mm}(b) \\
\includegraphics[scale=.15,clip,trim=60 60 60 60]{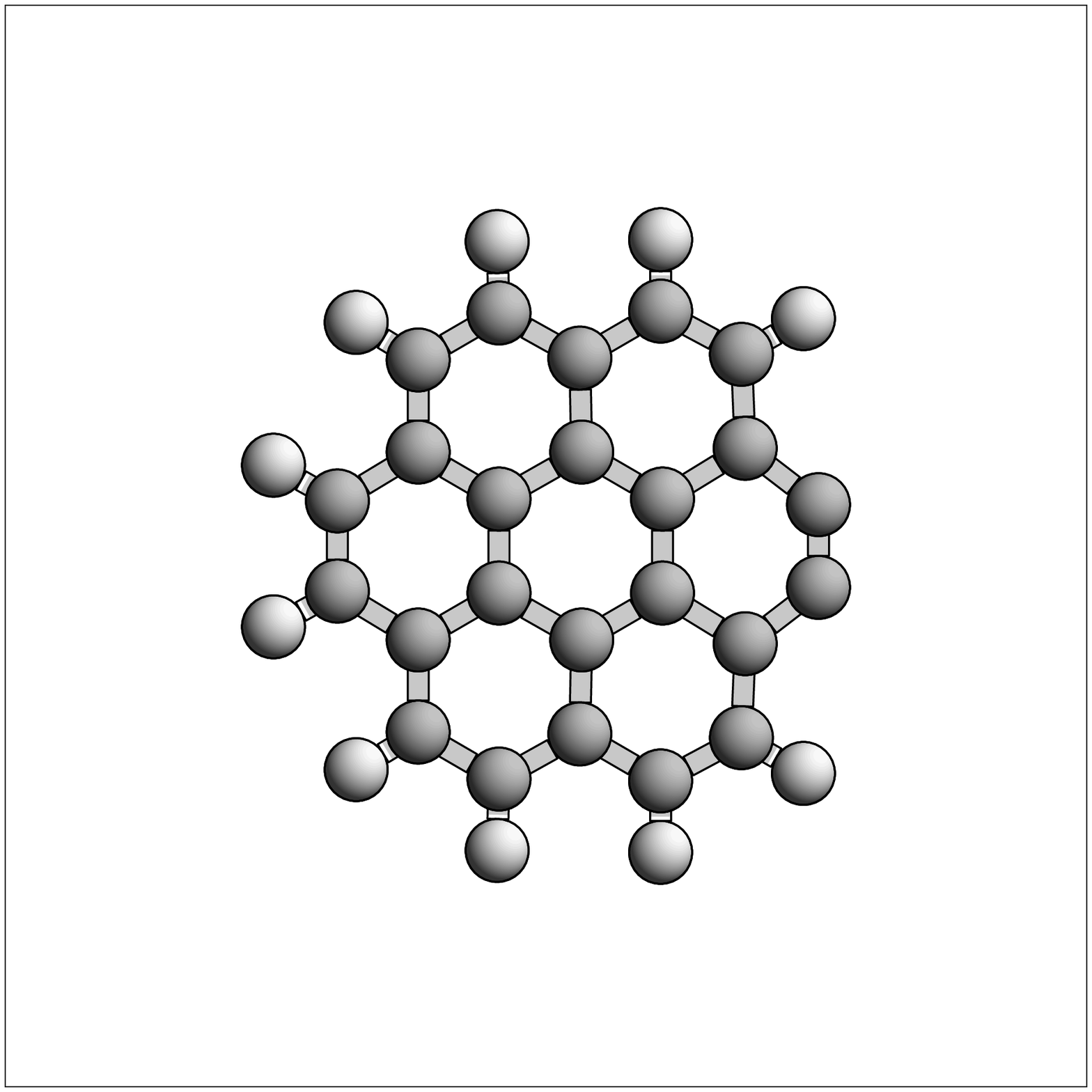}\hspace{-6mm}(c) \hspace{5mm}
\includegraphics[scale=.15,clip,trim=60 60 60 60]{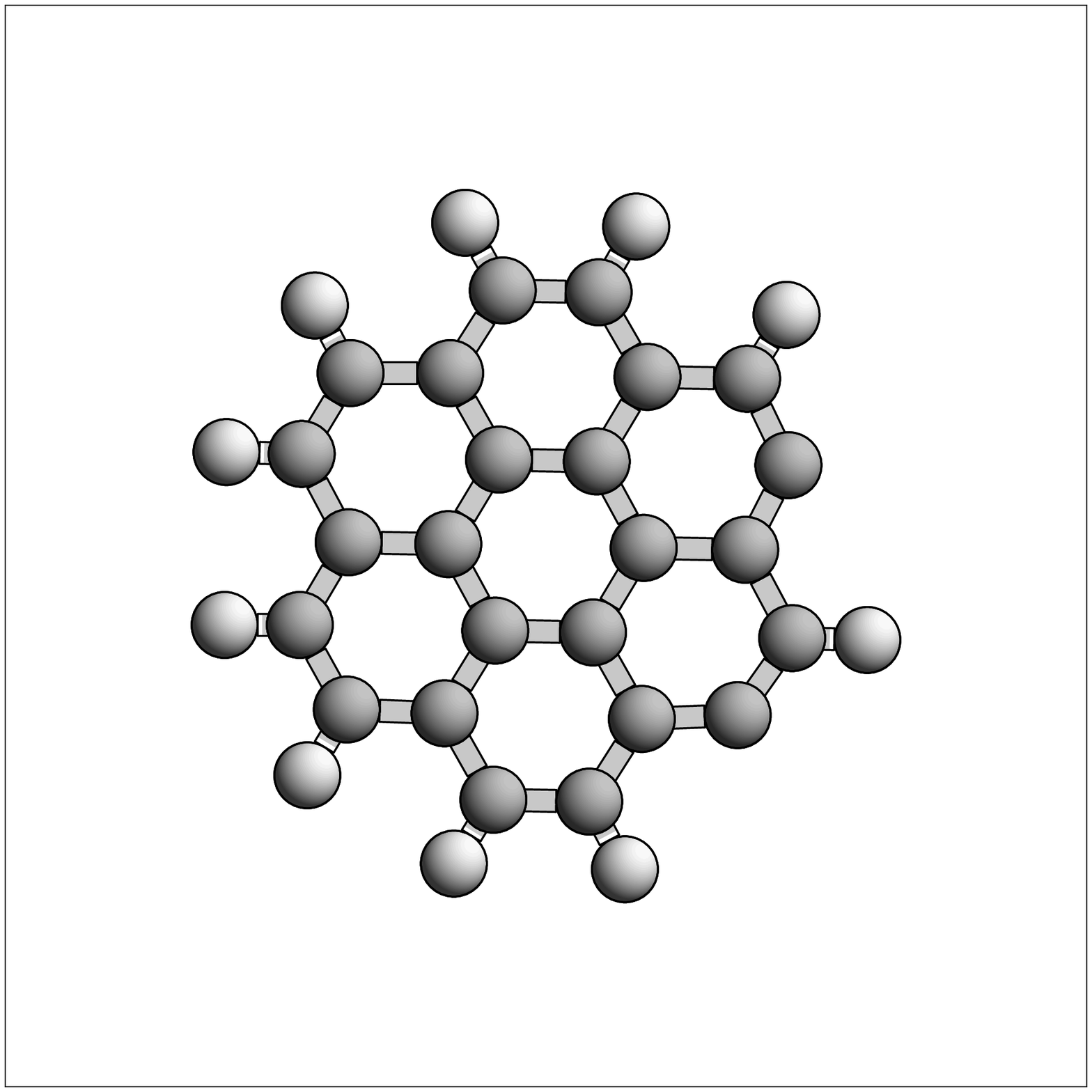}\hspace{-6mm}(d) \\ 
\includegraphics[scale=.15,clip,trim=60 60 60 60]{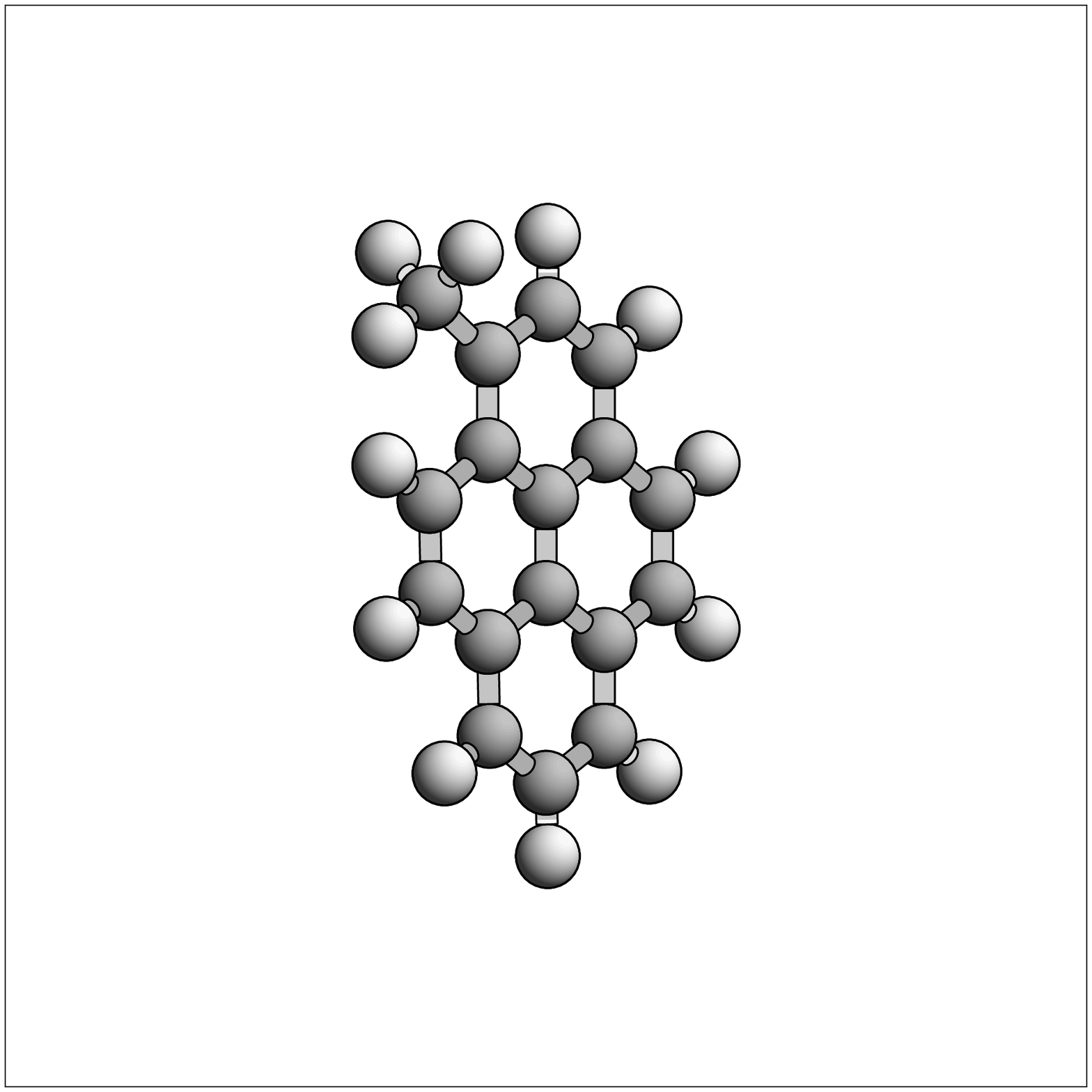}\hspace{-7mm}\includegraphics[scale=.15,clip,trim=60 60 60 60]{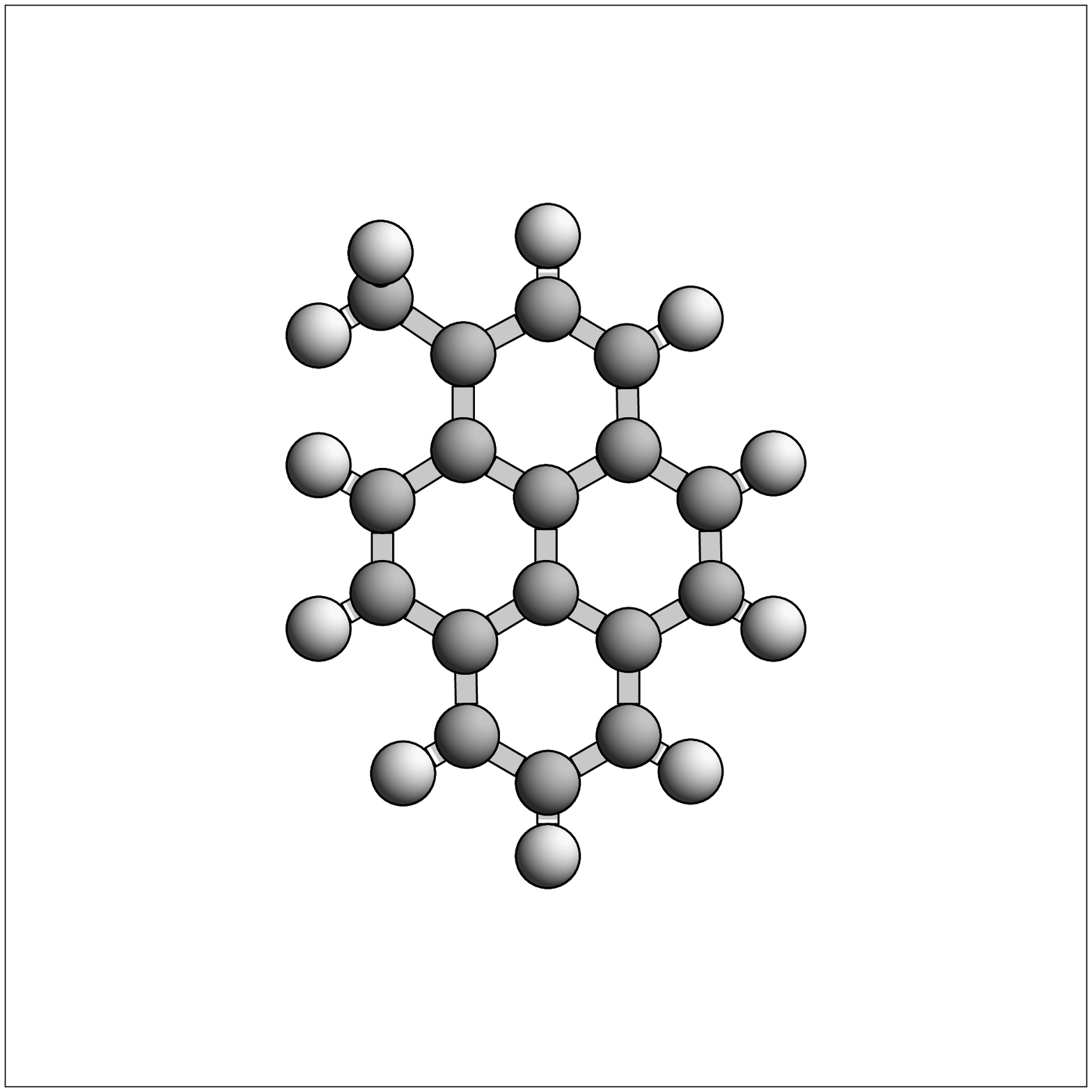}\hspace{-8mm}(e)\\
\includegraphics[scale=.15,clip,trim=60 60 60 60]{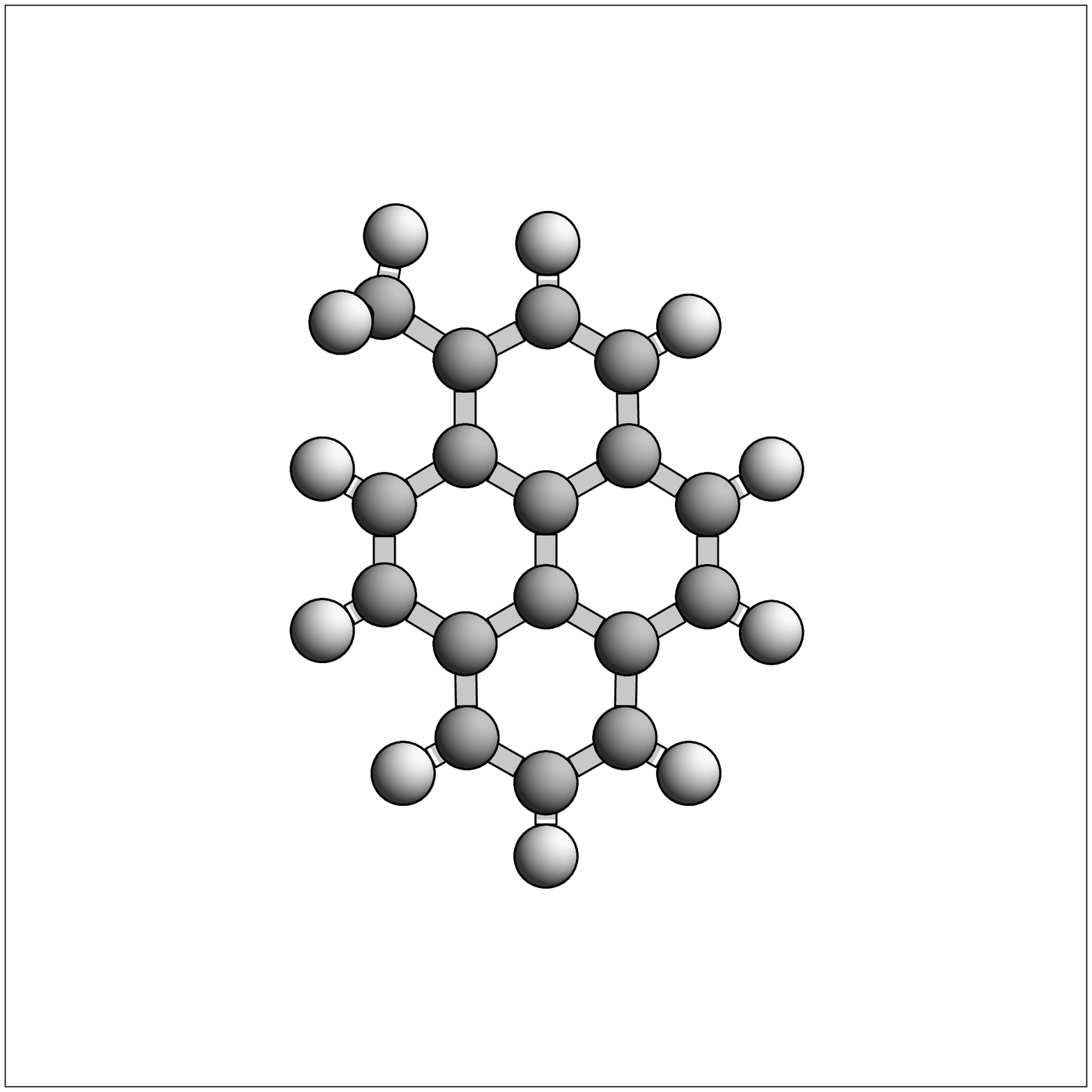}\hspace{-7mm}\includegraphics[scale=.15,clip,trim=60 60 60 60]{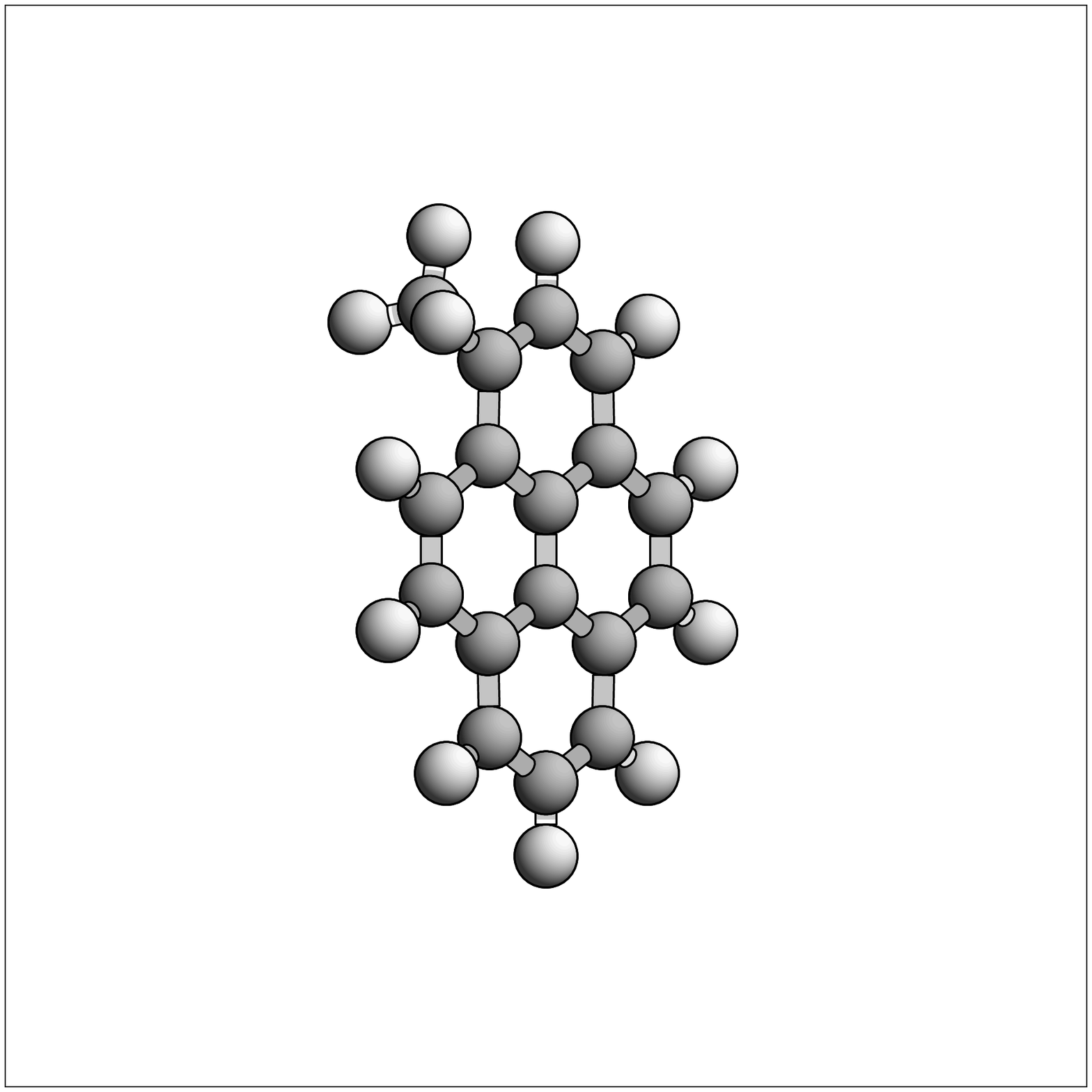}\hspace{-8mm}(f)
\caption{Structures of the studied molecules: (a)~pyrene cation; (b)~coronene cation; 
(c)~doubly dehydrogenated coronene cation with adjacent H vacancies; 
(d)~doubly dehydrogenated coronene cation in one of the configurations 
with non\textendash adjacent H vacancies; (e) and (f) the two stable 
conformations of the 1\textendash methylpyrene cation, each viewed from two 
different angles to better show the 3\textendash D configuration of the methyl
group.}
\label{molgeometries}
\end{center}
\end{figure}

\subsection*{The pyrene cation (${C_{16}H_{10}}^+$)}

\begin{figure}[h]
\begin{center}
\includegraphics[width=9cm]{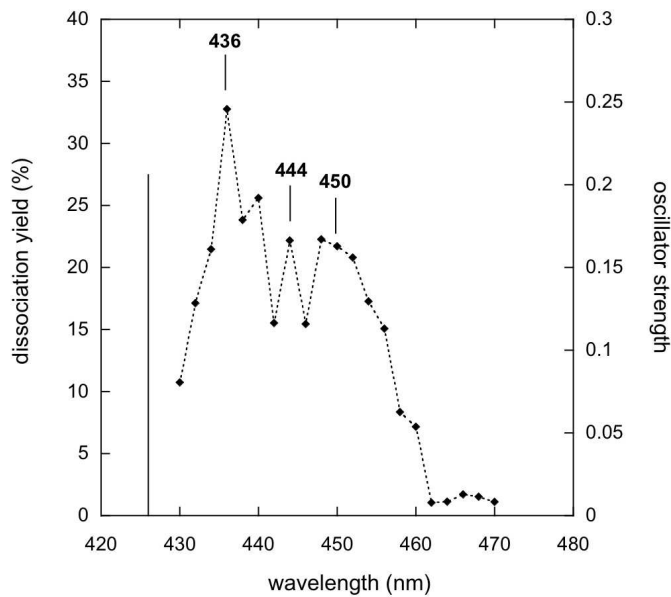}
\caption{Measured MPD spectrum of gas-phase pyrene cation isolated in the ICR cell of PIRENEA. Computed transitions are represented by vertical bars.}
\label{pyrene}
\end{center}
\end{figure}

The spectrum of the pyrene cation (see (a) in Figure~\ref{molgeometries}) is shown in Figure~\ref{pyrene}. 
We have recorded fragmentation after one laser pulse with a laser energy of 6~mJ. 
We can distinguish in the spectrum a main band at about 436~nm followed by two 
other peaks at about 444 and 450 nm. The band measured at 436 nm compares 
quite well with measurements in the gas phase made by Biennier et al.~\cite{Bien04} 
who assigned this feature to the (0-0) vibrational band of the 
$D_5 \longleftarrow D_0$ electronic transition. Our spectrum is also in 
good agreement with neon-matrix data taken by Salama and Allamandola.~\cite{Sal92}, 
(cf. Table~\ref{table}). The blue shift of about 3.5 nm that we measure relative 
to the matrix spectrum is consistent with the shifts measured by Biennier
et al.~\cite{Bien04} and with the predicted limits of the neon matrix to 
gas-phase shift ~\cite{Rom99,Sal99}.
Our theoretical calculations predict, in the considered spectral range, one electronic 
transition at 426~nm, with an oscillator strength \emph{f} = 0.206. This value is in line 
with previous calculations made by Hirata et al.~\cite{hirata99} and is in very good 
agreement with our experimental data ($\Delta \lambda $ = 10~nm or $\Delta$E = 0.07 eV 
with respect to our measured position). The two features observed at 444 and 450 nm are 
probably a vibronic progression of the observed electronic transition, since our calculations 
do not predict any other suitable electronic transition in this energy range. The fact that they 
are on the red side of the 0\textendash 0 band probably means that they are hot bands of the 
same electronic transition, in which the absorption takes place from a vibrationally excited state 
of the $D_0$ electronic state to a vibrationally less excited state of the $D_5$ electronic state. This indicates that the studied ions were initially carrying some internal energy transferred 
to them during the collisions with the buffer gas.
More vibronic structure can be present on the blue side of the band origin but we have not 
detected it because it is outside the probed spectral range.

\subsection*{The 1-methylpyrene cation (${CH_{3}-C_{16}H_{9}}^+$)}
\begin{figure}[h]
\begin{center}
\includegraphics*[width=9cm]{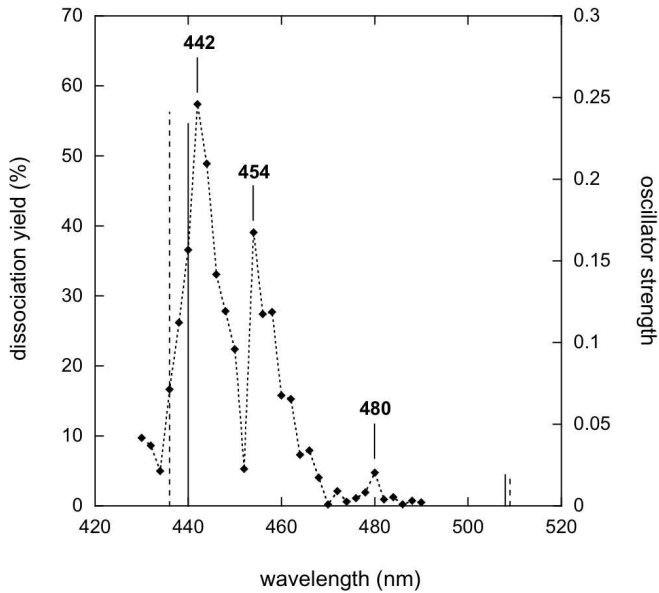}
\caption{Same as Figure \ref{pyrene} for 1-methylpyrene cation. The electronic transitions calculated for the two different isomers (e) and (f) (see Figure \ref{molgeometries}) are represented in continuous and dashed lines respectively.}
\label{methyl}
\end{center}
\end{figure}

Figure~\ref{methyl} shows the spectrum we measured for the 1\textendash methylpyrene cation. We performed experiments with one laser pulse, with a laser energy of 3 mJ. We can compare our results with the spectrum of UV-irradiated 1-methylpyrene isolated in a neon matrix recorded by L\'{e}ger et al.~\cite{Leg95} (cf. Table~\ref{table}). The authors measured four bands respectively at 418, 444, 456 and 482.5 nm. { Considering that some of the species present in the matrix could be PAH radicals that have lost an H atom, the authors incorporated atomic hydrogen (H$^0$) in the matrix and examined the evolution of the relative intensities of the bands in order to determine to which species they belong.} They therefore tentatively attributed to the 1\textendash methylene\textendash pyrene cation, presumably formed inside the matrix after irradiation, all the above-mentioned bands apart from the one at 456 nm ascribed to 1\textendash methylpyrene cation. In our spectrum we observe three bands at the following positions: 442, 454 and 480 nm. Taking into account the precision of our technique and the matrix to gas-phase shift, all these bands are likely to correspond, in position, to the bands measured in the matrix at 444, 456 and 482.5 nm. Band profiles and relative intensities also compare well with matrix data (cf.~\cite{Leg95}). The gas-phase spectrum of 1-methylpyrene cation has been previously measured by Tan \& Salama \cite{Tan}. The authors assigned to the $D_5 \longleftarrow D_0$ electronic transition of the molecule a spectral feature measured at $\sim$441 nm estimating for it an oscillator strength ${f}\sim 0.2$. This value matches well with our theoretical calculations which predict an electronic transition with the same oscillator strength at $\sim$440~nm (cf. Table~\ref{table}) and with our experimental measurement. Unfortunately, the authors only report on a narrow spectral window around the $\sim$441~nm feature, saying nothing about the other features. Given the technique we use here, the additional features we see at 454 and 480~nm cannot be ascribed to the 1\textendash methylene\textendash pyrene cation, but only to the 1\textendash methylpyrene cation. TD\textendash DFT calculations, on the other hand, only predict one additional transition in that spectral region, namely the, $D_4 \longleftarrow D_0$, at a wavelength of 507 nm  with an oscillator strength of  ${f} = 1.9 \times 10^{-2}$. This band probably corresponds to the transition we observed at 480 nm. The band at 454 nm in our spectrum could, in principle, be due to hot bands, as discussed above for pyrene, but then it should not be present in the cold neon-matrix spectra. However, our calculations show that there are two stable configurations for the methyl group, differing by a 180$^o$ rotation of the torsional angle (see (e) and (f) in Fig.~\ref{molgeometries}). The conformation (f) in Fig.~\ref{molgeometries} results to be marginally more stable by $\sim$0.04~eV, but such a difference is so small that it is within the error of DFT. A more accurate quantum chemistry method (e.g. coupled cluster) would be needed to reliably establish which is the most stable conformation, but that is out of the scope of the present paper and is not needed to interpret the experimental spectrum. With such a small energy difference, both conformers can be expected to be simultanously present both in our experiment and in cold matrix spectra. Their electronic transitions differ slightly in position and intensity, so that all well\textendash resolved bands are \emph{expected} to be double. We therefore attribute the two bands at 442 and 454 nm to the same $D_5 \longleftarrow D_0$ transition, predicted to be at 436~nm with \emph{f} = 0.241 for the (f) isomer and at 440~nm with \emph{f} = 0.234 for the (e) isomer.
{Considering the precision of our calculations this assignment is compatible with the experimentally observed splitting of 12 nm.}

\subsection*{The coronene cation (${C_{24}H_{12}}^+$) and its doubly dehydrogenated derivative (${C_{24}H_{10}}^+$)}

\begin{figure}[h]\begin{center}
\includegraphics[width=9cm]{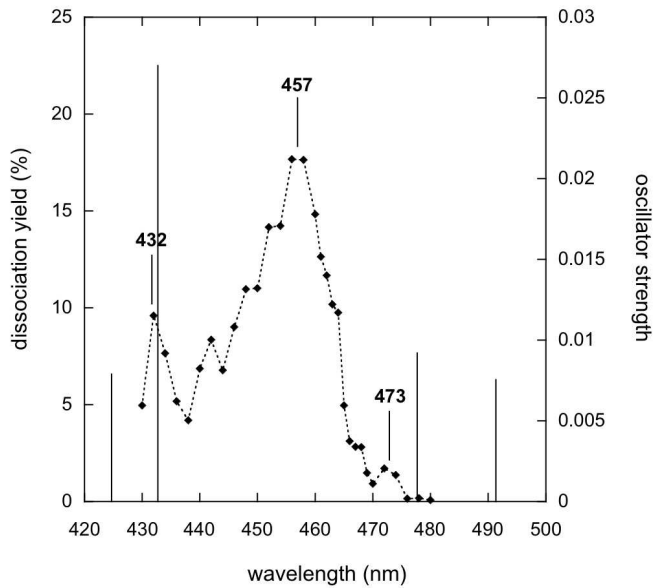}
\caption{Same as Figure \ref{pyrene} for coronene cation. Only the electronic transitions calculated with D$_{2h}$ symmetry are shown.}
\label{coronene}
\end{center}
\end{figure}

\begin{figure}[h]
\begin{center}
\includegraphics[width=9cm]{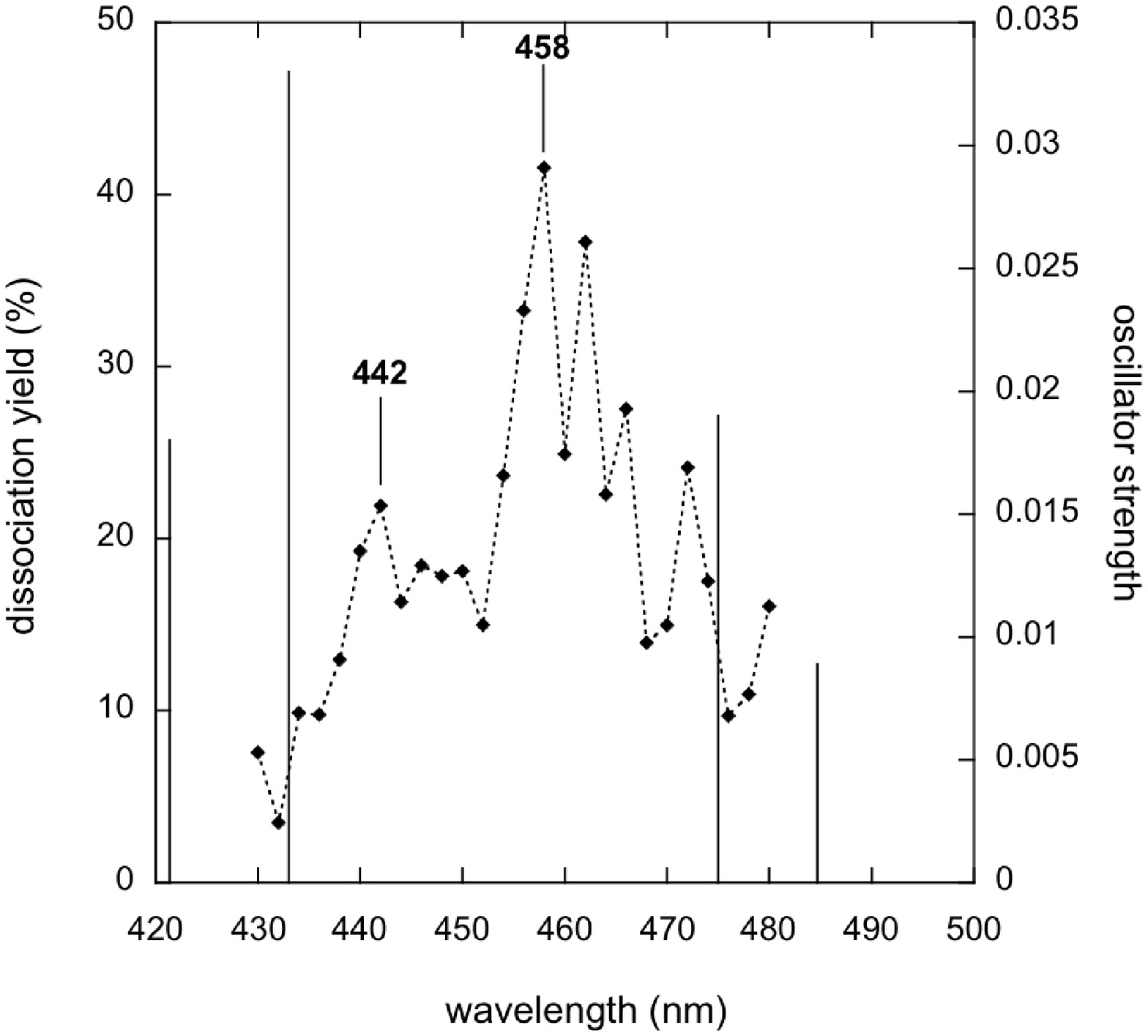}
\caption{Same as Figure \ref{pyrene} for doubly-dehydrogenated coronene cation.}
\label{dehydro}
\end{center}
\end{figure}

The spectra of gas-phase coronene cation (see (b) in Figure \ref{molgeometries}) and of one of its dehydrogenated derivatives are reported here for the first time (see Figures~\ref{coronene} and \ref{dehydro}). Dissociation of coronene ions was performed with one OPO laser shot and with a laser energy of 14 mJ. Our spectrum presents a broad feature at about 457 nm with some vibronic structure on the blue side of the band (cf. Figure \ref{coronene}). Identification of this band through comparison with theoretical calculations is not trivial for several reasons: with our experimental technique we have no information on the absolute intensity of the transition and the small spectral range analysed makes it difficult to unambiguously assign bands. Moreover, the ground state of the coronene cation slightly breaks the $D_{6h}$ symmetry due to the Jahn\textendash Teller effect. This affects excited states in a different way (some states are not affected at all, depending on symmetry of the state) and is relatively prone to numerical errors, since Jahn\textendash Teller minima are usually very shallow. Therefore, computing just TD\textendash DFT vertical transitions from the optimised (symmetry\textendash broken) geometry of the ground state is a slightly worse approximation than in other cases. Symmetry\textendash broken states, in principle, should be properly treated by taking into account the dynamic interaction of the two splitted electronic configurations, since they are separated by a very small energy barrier. This is a level of theory far beyond the scope of the present paper. We can however estimate the expected error range of the calculations by comparing the same vertical electronic transitions, as computed by enforcing $D_{6h}$ symmetry, with those calculated in the geometry obtained optimising the ground state after relaxing the symmetry constraint (which results in $D_{2h}$ symmetry due to Jahn\textendash Teller distortion). Moreover, to ease the problem of the small spectral window, we can first compare calculations with experimental matrix data taken from Ehrenfreund et al. \cite{Ehre92} that include all the visible spectral range, and systematically assign the transitions on it.
TD\textendash DFT predicts four electronic transitions between 400 and 500 nm, respectively at 425, 433, 478, and 491~nm calculated with D$_{2h}$ symmetry and at 419.8, 420.3, 470, and 505~nm calculated with D$_{6h}$ symmetry (cf. Table \ref{table}). In the matrix spectrum published by Ehrenfreund et al. \cite{Ehre92}, the most intense band in this interval is detected at 459~nm and matches the one we detect at 457~nm. We tentatively assign it to the $D_{10} \longleftarrow D_0$, calculated to be at 433~nm with an oscillator strength $f = 2.7 \times 10^{-2}$. The partially resolved structure on the blue side, which is visible both in our experiments and in the neon-matrix spectrum, is likely to be a vibronic sequence of the same electronic transition. The band that we detect at 432~nm would then be the $D_{11} \longleftarrow D_0$, calculated to be at 425~nm with an oscillator strength $f = 8.0 \times 10^{-3}$. We further detect another weak band at 473~nm, which is also visible in the neon-matrix spectrum \cite{Ehre92}, even if unlisted in this paper, which would be the $D_9 \longleftarrow D_0$, calculated to be at 478~nm with an oscillator strength $f = 9.2 \times 10^{-3}$. Finally, the band detected at 502~nm in the neon-matrix spectrum would then be the $D_5 \longleftarrow D_0$, calculated to be at 491~nm with an oscillator strength $f = 7.6 \times 10^{-3}$, even if at least part of this band could be due to a coincident very strong band in benzo(g,h,i)perylene~\cite{Sal99}, which could be formed in the matrix experiment by fragmentation of coronene. 
The comparison between the two theoretical calculations, obtained with and without the D$_{6h}$ symmetry constraint, and corresponding to very close geometries, show the sensitivity of band positions and intensities to the conformation of the molecule. This elucidates the difficulty in interpreting the intensities in MPD spectra in terms of absorption intensities: since several photons are needed to achieve photodissociation, all but the first are absorbed by a hot molecule, in which the band position and intensities may be shifted by variable amounts. This means that a band whose position is more sensitive to geometry changes will appear weaker on MPD spectra, since the absorption of the subsequent photons will be less likely. Conversely, a band relatively insensitive to small geometry changes will appear stronger.

\begin{table*}[htbp]
\begin{center}      
\caption{Positions of the bands (expressed in nm) as measured in the PIRENEA experiment in the wavelength range 430-480~nm 
for the four gas-phase species considered. Band origins are identified through the comparison with the electronic transitions and the corresponding oscillator strengths (within parentheses) as computed using the split valence polarization (SVP) basis set in conjunction 
with the BP86 exchange-correlation functional. For comparison we list also previous experimental and theoretical data.}
\label{table}
\begin{tabular}{cccccc}
\hline
&\multicolumn{2}{c}{Experiment} & \multicolumn{3}{c}{TD-DFT}\\
Transition & PIRENEA & Previous & \multicolumn{2}{c}{BP/SVP} &Previous\\
\hline 
\multicolumn{6}{c}{Pyrene$^+$ (C$_{16}$H$_{10}^{+}$)} \\
\hline
$D_5 \longleftarrow D_0$ 
& 436 & 439.5$^c$, 436.2$^d$ &  \multicolumn{2}{c}{426(0.206)} & 419(0.291)$^b$\\
$D_4 \longleftarrow D_0$
 & & 486.8$^a$ &   \multicolumn{2}{c}{508(0.012)} & 498(0.017)$^b$\\
\hline
\multicolumn{6}{c}{1-Methylpyrene$^+$ (CH$_3$-C$_{16}$H$_{9}^{+}$)} \\
\hline
& & & isomer (e) & isomer (f) & \\
$D_5 \longleftarrow D_0$ 
& 442 & 444.0$^e$, $\sim$441$^f$ & & 436(0.241) & \\
& 454 & 456.0$^e$& 440(0.234) & & 440(0.233)$^g$\\
$D_4 \longleftarrow D_0$
& 480 & 482.5$^e$& 508(0.019) & 509(0.017) & 507(0.019)$^g$\\ 
\hline
\multicolumn{6}{c}{Coronene$^+$ (C$_{24}$H$_{12}^{+}$)} \\
\hline
& & & D$_{2h}$& D$_{6h}$& \\
$D_{12} \longleftarrow D_0$ 
& & & & 419.8(0.007) & \\
$D_{11} \longleftarrow D_0$ 
& 432 & & 425(0.008) & 420.3(0.007) & \\
$D_{10} \longleftarrow D_0$
& 457 & 459.0$^h$ & 433(0.030) & & 427(0.030)$^i$\\
$D_{9} \longleftarrow D_0$ 
& & & & 470(0.008) & \\
$D_{7} \longleftarrow D_0$ 
& 473 &  & 478(0.009)& & \\
$D_{6} \longleftarrow D_0$ 
& & & & 505(0.002) & \\
$D_{5} \longleftarrow D_0$ 
& & 502.0$^h$ & 491(0.008)& &\\
\hline
\multicolumn{6}{c}{Doubly dehydro-coronene$^+$ (C$_{24}$H$_{10}^{+}$)} \\
\hline
$D_{17} \longleftarrow D_0$
& 442 & & \multicolumn{2}{c}{421(0.018)} & \\
$D_{16} \longleftarrow D_0$
& 458 & & \multicolumn{2}{c}{433(0.033)}&\\
$D_{11} \longleftarrow D_0$
& &  & \multicolumn{2}{c}{475(0.019)}& \\
$D_{9} \longleftarrow D_0$
& & & \multicolumn{2}{c}{485(0.009)}& \\
\hline
\end{tabular}
\end{center}
$^a$ Argon matrix \cite{vala}.\\
$^b$ TD-DFT calculations at the BLYP/{6-31G$^{**}$} level
\cite{hirata99}.\\
$^c$ Neon matrix \cite{Sal92}.\\
$^d$ Multiplex integrated cavity output spectroscopy~\cite{Bien04}.\\
$^e$ Neon matrix \cite{Leg95}.\\
$^f$ Cavity ring-down spectroscopy~\cite{Tan}.\\
$^g$ TD-DFT calculations at the BP86/\mbox{SVP} level \cite{Tan}.\\
$^h$ Neon matrix \cite{Ehre92}. \\
$^i$ TD-DFT calculations at the BLYP/\mbox{6-31G$^{*}$} level
\cite{wei03}.\\
\end{table*}

We also measured, for the first time, the gas-phase spectrum of the dehydrogenated derivative of coronene ${C_{24}H_{10}}^+$. This was obtained in the ICR cell by photofragmentation of the coronene parent ion by UV irradiation with a Xe arc lamp. We mass\textendash selected the fragments and performed MPD spectroscopy. ${C_{24}H_{10}}^+$ thus produced can, in principle, have six nonequivalent isomers corresponding to different distances between the photoejected H atoms, which cannot be distinguished \emph{a priori} in our setup. However previous photodissociation experiments performed on the coronene cation in PIRENEA, provide strong evidence that only the isomer with adjacent H atoms missing has to be considered~\cite{Joblin10}. This is further confirmed by calculations on the binding energies of the isomers which show that this isomer is, by far, the most stable one (about 1.2 eV lower in energy). MPD was performed with one laser pulse at 12 mJ. In the ${C_{24}H_{10}}^+$ spectrum (cf. Figure~\ref{dehydro}) two closely spaced electronic transitions stand out, at $\sim$ 442 and 458 nm. We observe also what could be interpreted as a vibronic progression on the red side of the 458~nm band. {As discussed in the case of the pyrene cation, this indicates that the ion cloud is retaining some residual excitation energy due to the ejection process, performed in this case to eject the parent ion dominant species.} The strongest calculated transition in this spectral range is predicted to be the $D_{16} \longleftarrow D_0$ at 433~nm, with $f = 3.3 \times 10^{-2}$. This transition is therefore the most likely assignment for the strongest peak we observe at 458~nm. The band we observe at 442~nm can then be associated to the $D_{17} \longleftarrow D_0$ transition predicted to be at 421 nm with $f = 1.8 \times 10^{-2}$ (see Table~\ref{table}).

\section{Modelling the photophysics of PAH cations}
\label{Model} 
The dissociation yield that is measured in our experiments depends on the absorption cross-section, $\sigma_{abs}$, of the studied species and on the local intensity of the radiation field over the ion cloud. In order to retrieve the value of $\sigma_{abs}$ it is necessary to interpret the experimental data using a model that describes the photophysics of the ions.
To achieve this goal we have used a kinetic Monte Carlo code~\cite{Rap,Job} that provides a description of the photoabsorption and dissociation processes and the time evolution of the internal energy of the considered species as a function of the flux density of photons. The method has been initially described by~\cite{Boissel}.
In our experimental conditions, involving low values of the laser fluence, multiphoton events are negligible. However, during a laser pulse ($t_{pulse}$ = 5 ns), the ion has enough time to absorb sequential photons leading to an increase of its internal energy. If this energy is above the dissociation threshold the ion may fragment. 

The absorption rate constant, $k_{abs}$, can be expressed by the following equation
\begin{equation}
k_{abs}  =  \frac {\sigma_{abs}} {S} \phi = \frac {\sigma_{abs}} {S} \frac {\lambda}{hc} \frac {E_{laser}}{t_{pulse}}
\end{equation}
where $\sigma_{abs}$ is the absorption cross-section of the electronic state of the molecule averaged over the spectral profile of the laser, S the laser spot surface, $\phi$ is the photon flux (number of photons per unit time), $\lambda$ is the wavelength of the excitation photon and $E_{laser}$ and $t_{pulse}$ are, respectively, the laser pulse energy and the pulse duration. 

The fragmentation rate $k_d$ as a function of the internal energy U, is given by
\begin{equation}
k_{d}(U) = A_d \frac {\rho (U - E_d)} {\rho (U)} 
\end{equation}
where $\rho$ is the density of states of the parent ion, $E_d$ is the binding energy which is taken to be 4.8 eV for the loss of an H atom~\cite{Pino,Jolibois} and $A_d$ is a preexponential factor. {This equation was derived in~\cite{Boissel} and is a simplified version of the phase-space theory used in ~\cite{Pino}}.
The density of states is calculated using the list of modes for the ion and applying the Beyer-Swinehart algorithm~\cite{Stein}.
The value of $A_d$ was calculated using the results of Jochims et al.~\cite{Jochims}: $k_d = 10^{4}~s^{-1}$ at U = 9.06 and 12.05 eV for ${C_{16}H_{10}}^+$ and ${C_{24}H_{12}}^+$respectively.

\begin{figure}[!h]
\begin{center}
\begin{minipage}{8cm}
\includegraphics[width=8cm]{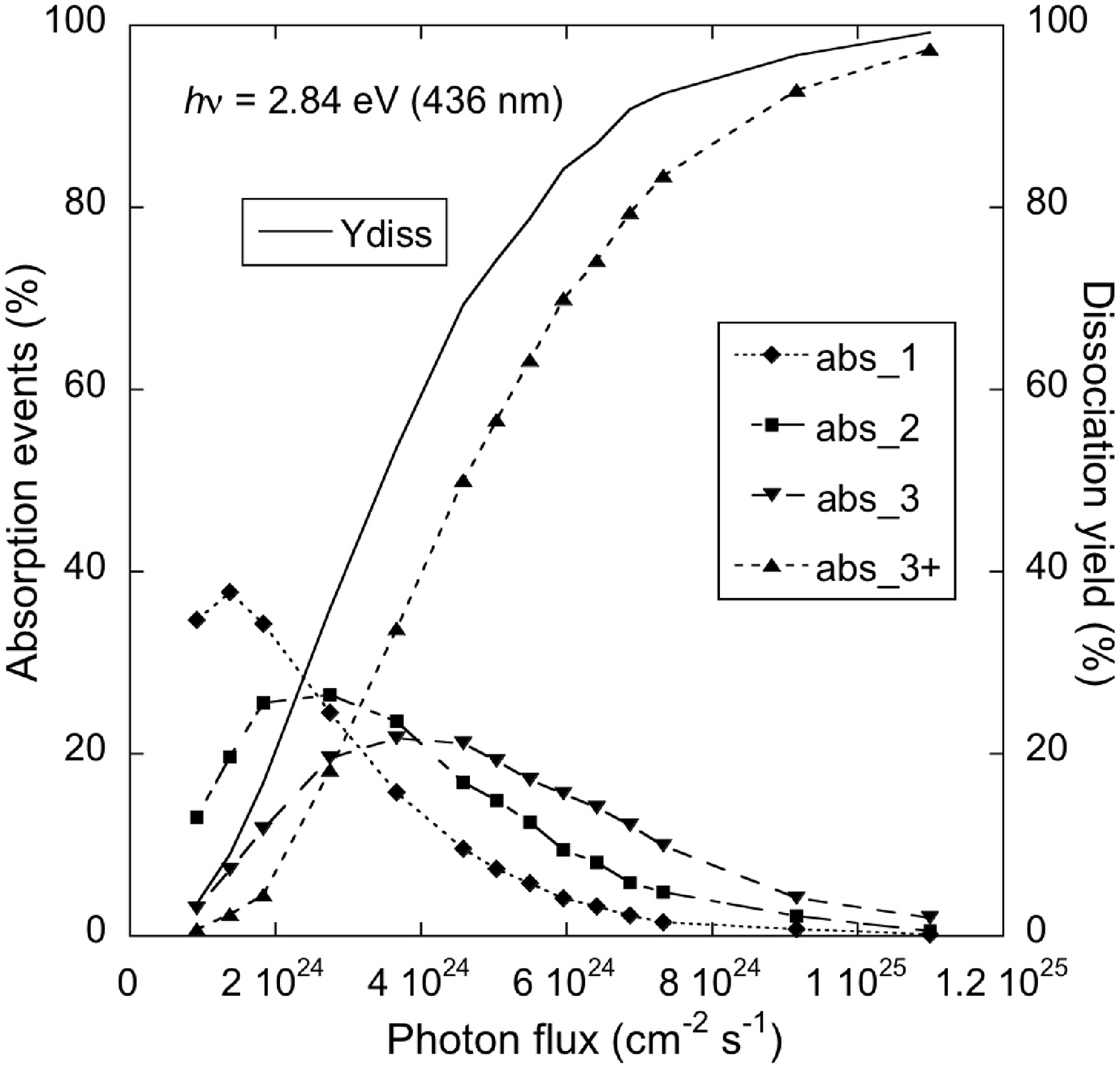}
\end{minipage}
\begin{minipage}{8cm}
\includegraphics[width=8cm]{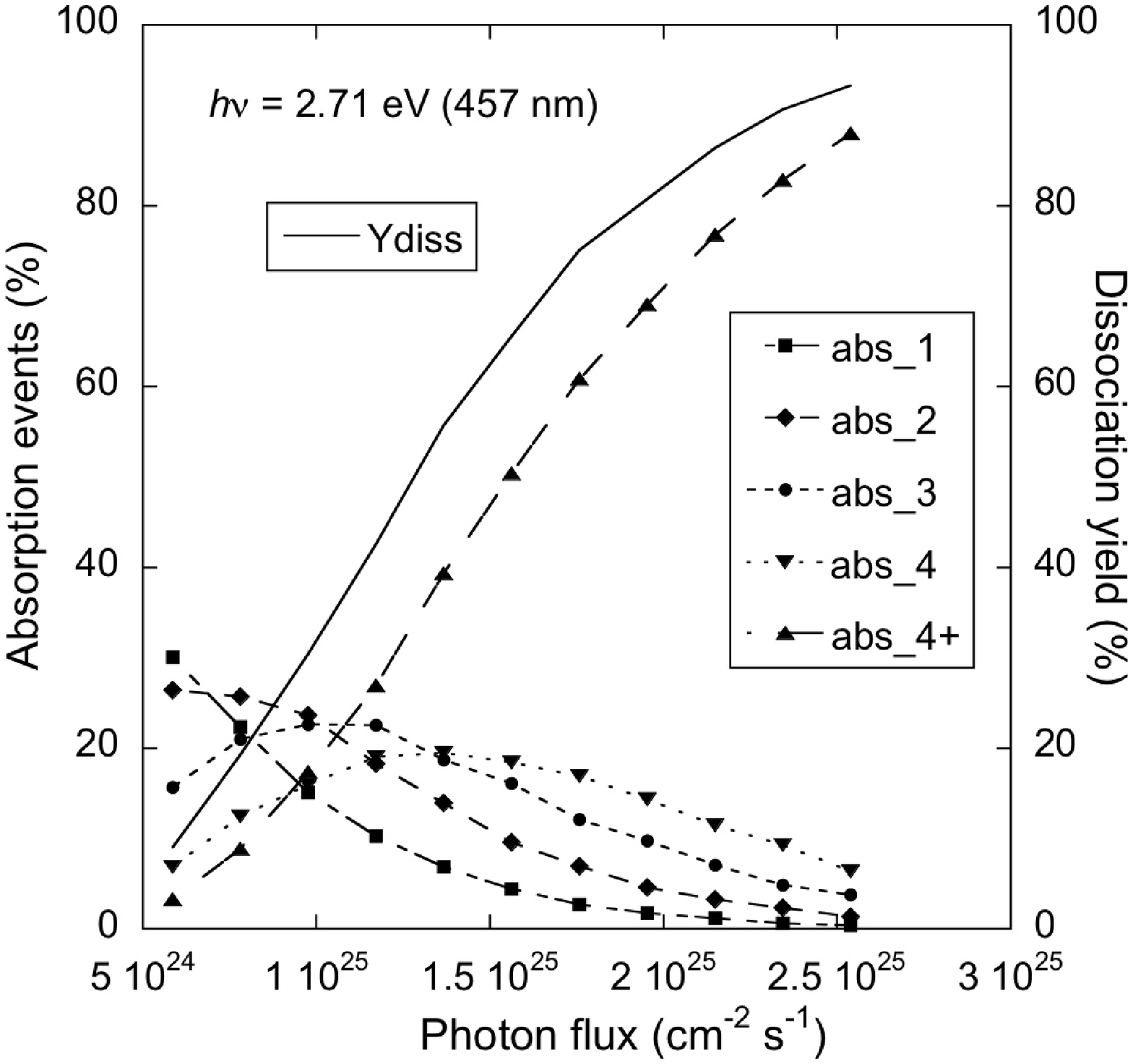}
\end{minipage}
\caption{Photoabsorption events (abs\_n is the fraction of events with n absorbed photons, abs\_n+ with more than n photons) and total dissociation yield (Ydiss) calculated with the model as a function of the photon flux. The upper panel shows the results obtained for the pyrene cation for $\sigma_{abs} = 1.6 \times 10^{-16} cm^2$ while the bottom panel the results obtained for the coronene cation for $\sigma_{abs} = 0.6 \times 10^{-16} cm^2$.}
\label{diss_model}
\end{center} 
\end{figure}

We have observed that the experimental conditions strongly influence the measurement of the dissociation yield. Two additional parameters have to be considered when modelling the data: the overlap between the laser spot and the ion cloud, which can change due to possible variations of the ion cloud size, shape and position, and the photon local density in the laser spot, which in turn depends on the laser beam profile. 
The ion cloud spatial distribution and position inside the cell depend on the details of ion formation and injection into the trap, which can to some extent vary from shot to shot. This affects the overlap between the laser spot and the ion cloud, and consequently the measured dissociation yield of the species. This is taken into account in our measurements by averaging, at each wavelength, the recorded dissociation yield over several spectra.
The second effect we have to consider is the variation of the local photon flux (number of photons per unit time per unit surface) in the laser spot. This parameter plays an important role on dissociation, especially in the case of molecules that need to absorb a large number of photons to dissociate (e.g. coronene cation). 
The OPO beam intensity follows a gaussian profile over an elliptical surface with semi-axis of, respectively, 3 and 2.55 mm (values obtained from the FWHM of the x and y beam profiles).
The beam profile gives us the spatial distribution of the laser intensity over the laser spot surface from which we can determine spatial distribution of the photon flux in the laser spot. Using the Monte Carlo model, we were able to derive the number of photons absorbed per pulse and the corresponding dissociation counts for different values of the absorption cross-section. Some of the results are presented in Figure~\ref{diss_model}. As can be seen, the dissociation is dominated, in the case of the pyrene cation, by the ions that have absorbed at least three photons of 2.84 eV ($\lambda$ = 436 nm) while for the coronene cation, the absorption of four photons of 2.71 eV ($\lambda$ = 457 nm) is required to efficiently dissociate.
Some recent studies on the photofragmentation of fluorene cation using sequential multiphoton absorption have shown that the absorption cross-section may vary with the number of absorbed photons~\cite{VanI,VanII}. We have considered this hypothesis in our calculations running simulations for the pyrene cation with a variable cross-section but we could not evidence such an effect from our measurements. 

\begin{figure}[!h]
\begin{center}
\begin{minipage}{8cm}
\includegraphics[width=8cm]{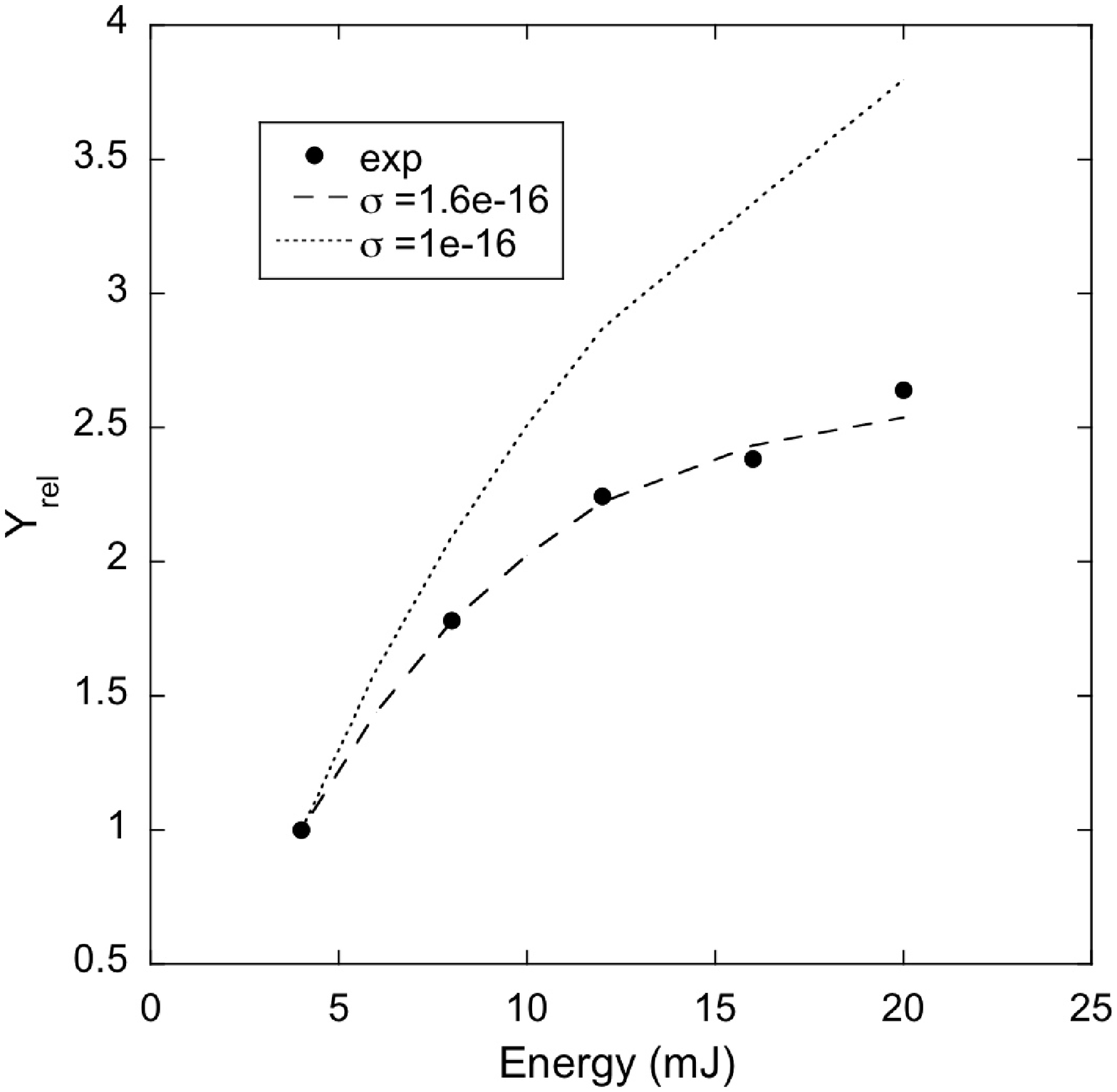}
\end{minipage}
\begin{minipage}{8cm}
\includegraphics[width=8cm]{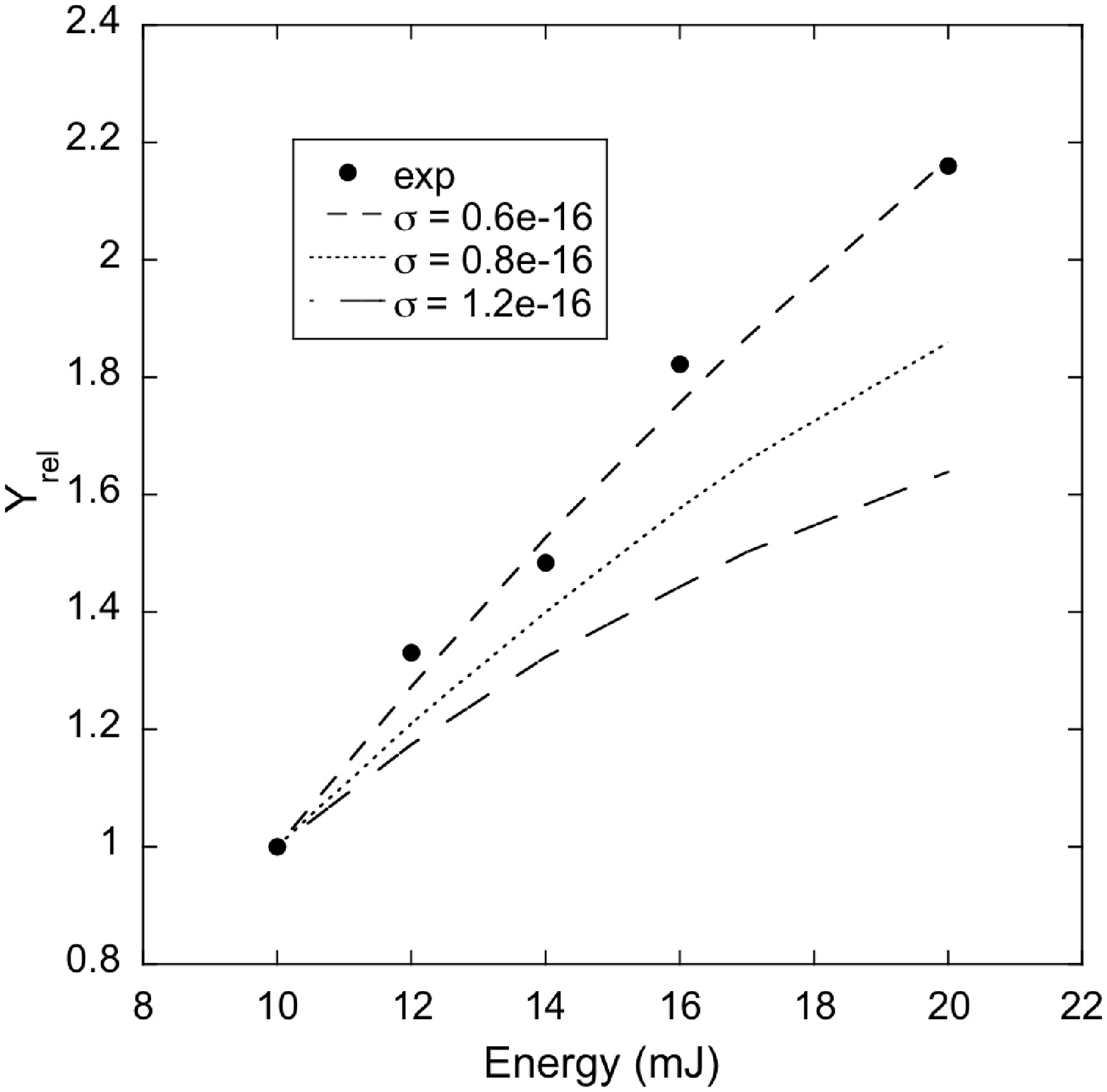}
\end{minipage}
\caption{Relative dissociation yields measured experimentally (dots) and normalised to the lowest value, compared with the results obtained from the model considering different values for the absorption cross-sections (lines), for pyrene cation (upper panel) and coronene cation (bottom panel).}
\label{Yrel}
\end{center}
\end{figure}

To derive the absorption cross-section from fitting of the experimental data with our model we have measured the dissociation yield as a function of the laser energy. We observed an increase of this dissociation yield with the energy up to a saturation value that is on average 50\% and 60\% for ${C_{16}H_{10}}^+$ and ${C_{24}H_{12}}^+$ respectively, and up to 80\% when considering individual shots. {The variation of the dissociation yield from one laser shot to another is related to the variation in the local photon flux in the laser spot, as mentioned above. The laser intensity decreases going from the centre of the spot to the edges, so ions located at the centre of the laser spot will dissociate more efficiently than ions intercepted by the edges of the spot. Considering that a dissociation yield up to 80\% can be attained on individual shots, we can suppose that the size of the ion cloud is of the same order of the laser beam or probably bigger.}

The calculated dissociation yield is obtained combining the results of the Monte Carlo simulations (Figure~\ref{diss_model}) and the spatial information on the photon flux density derived from measurements of the laser beam profile. Relative dissociation yields were used to obtain values that are independent of the ion cloud size.
The fitting of the experimental data allows then to constrain the absorption cross-section of the studied species. We derived a value of $1.6 \times 10^{-16} \mathrm{cm}^2$ for the pyrene cation at $\lambda$ = 436 nm and of $0.6 \times 10^{-16} \mathrm{cm}^2$ for the coronene cation at $\lambda$ = 457 nm (cf. Figure~\ref{Yrel}).
The value obtained for the pyrene cation results to be the same estimated by Biennier et al.~\cite{Bien04}. In the case of the coronene cation our data represent the first estimation of its cross-section coming from gas-phase experimental measurements. An absorption cross-section $\sigma_{abs} \simeq 0.9 \times 10^{-16}  \mathrm{cm}^2$ for the 459 nm band can be derived from the matrix data of Ehrenfreund et al.~\cite{Ehre92}. This value is very close to our result. All these results justify \emph{a posteriori} that our physical description, in which we consider only sequential absorption of photons and neglect true multiphoton events, is correct.

\section{Conclusions}
\label{Con}
We used a technique which enables us to indirectly study the visible absorption spectrum of non-volatile reactive ions isolated in a cold ICR cell. The target ions can be produced \emph{in situ} by photofragmenting a suitable parent species.
Since all species with a different m/z ratio are ejected before recording the spectrum, the technique is free from contaminants (with the exception of isomers) that are difficult to avoid by other methods. The technique has however the limitation of recording an action spectrum that does not provide direct information on the cross-section. 
Still the results we obtained are encouraging for several reasons: 
\begin{itemize}
\item Comparison with gas-phase spectra of cold PAHs have shown a very good agreement with the measured positions confirming the validity of our experimental method in obtaining the band positions in gas-phase. 
\item Gas-phase spectra of ${C_{24}H_{12}}^+$ and ${C_{24}H_{10}}^+$ have been reported here for the first time adding an important contribution to the spectroscopic study of such large non-volatile molecules.
\item Theoretical calculations and modelling have been used to complement the technique. Theoretical calculations have shown, in general, a good agreement with our experimental data and with previous measurements found in the literature. Moreover a kinetic Monte Carlo code has been used, in the case of the pyrene and coronene cations, to constrain the photophysics of the ions and, in particular, to estimate their absorption cross-section. The obtained values are in satisfactory agreement with previous experimental results. 
\end{itemize}
{Among the species studied here only two of them exhibited spectral features at a position close to the position of the 4430~\AA~DIB, the 1\textendash methylpyrene cation ($\lambda_{PIRENEA}$ = 442 nm) and the doubly dehydrogenated coronene cation ($\lambda_{PIRENEA}$ = 442 nm). For the first of them, ${CH_{3}-C_{16}H_{9}}^+$, measurements made on cold gas-phase ions by Tan and Salama~\cite{Tan} have already shown that the characteristics of the observed band are not correlated with those of the 4430~\AA~DIB.
Still, an interesting result we have obtained with our measurements is the identification of the simultaneous presence of two different structures for this species. This aspect could not be evidenced in matrix experiments in which the contribution of another species (the 1\textendash methylene\textendash pyrene cation) was invoked to interpret the spectrum.
In the case of the doubly dehydrogenated coronene cation, two electronic transitions are present in the considered spectral range, at 442 and 458 nm, the second one being much stronger. 
This automatically excludes this species as a possible carrier for the 4430~\AA~DIB}.
Perspectives for this work will include a better control of the ion temperature in the trap.
From the theoretical point of view, modelling of the vibrational dynamics of hot ions and a detailed TD\textendash DFT study of the dependence of band positions and intensities on vibrational coordinates will enable us to more accurately reconstruct the absorption spectrum from our MPD measurements.

\section*{Acknowledgements}

This work was supported by the European Research Training Network ``Molecular Universe'' (MRTN-CT- 2004-512302), the French National Program "Physique et Chimie du Milieu Interstellaire" (PCMI), the CNRS PICS 4260 and
the PPF of the Toulouse University "Mol\'ecules et Grains: du laboratoire \`a l'Univers".

\end{document}